\documentclass[conference]{IEEEtran}
\IEEEoverridecommandlockouts
\usepackage{cite}
\usepackage{amsmath,amssymb,amsfonts}
\usepackage{algorithmic}
\usepackage{graphicx}
\usepackage{textcomp}
\usepackage{xcolor}
\usepackage{xspace}
\usepackage{hyperref}
\usepackage{pifont}

\usepackage{cleveref}

\newcommand{\aieblas}{\textsc{aieblas}\xspace}
\newcommand{\blas}{\textsc{BLAS}\xspace}

\newcommand{\one}{\ding{192}\xspace}
\newcommand{\two}{\ding{193}\xspace}
\newcommand{\three}{\ding{194}\xspace}
\newcommand{\four}{\ding{195}\xspace}

\newcommand{\axpy}{\texttt{axpy}\xspace}
\newcommand{\gemv}{\texttt{gemv}\xspace}
\newcommand{\axpydot}{\texttt{axpydot}\xspace}

\begin{document}

\title{Developing a BLAS library for the AMD AI
Engine  \\ {\Large \textsc{Extended Abstract}}}

\author{\IEEEauthorblockN{Tristan Laan}
\IEEEauthorblockA{
\textit{Department of Computer Science} \\
\textit{Vrije Universiteit Amsterdam}\\
t.laan2@student.vu.nl}
\and
\IEEEauthorblockN{Tiziano De Matteis}
\IEEEauthorblockA{
\textit{Department of Computer Science} \\
\textit{Vrije Universiteit Amsterdam}\\
t.de.matteis@vu.nl}
}

\maketitle

\begin{abstract}

Spatial (dataflow) computer architectures can mitigate the control and performance overhead of classical von Neumann architectures such as traditional CPUs.  
Driven by the popularity of Machine Learning (ML) workloads, spatial devices are being marketed as ML inference accelerators. Despite providing a rich software ecosystem for ML practitioners, their adoption in other scientific domains is hindered by the steep learning curve and lack of reusable software, which makes them inaccessible to non-experts. 
We present our ongoing project \aieblas, an open-source, expandable implementation of Basic Linear Algebra Routines (BLAS) for the AMD AI Engine. Numerical routines are designed to be easily reusable, customized, and composed in dataflow programs, leveraging the characteristics of the targeted device without requiring the user to deeply understand the underlying hardware and programming model.
\end{abstract}


\section{Introduction}

The end of Dennard scaling and Moore’s law, along with the growing computational needs of application domains such as machine learning, are compelling industry and researchers to rethink classical computer organization. One promising alternative is constituted by Spatial Architectures, which, moving away from the von Neumann model, aim to make more efficient use of the available transistors and chip space.

Spatial architectures are today marketed mainly as Machine Learning (ML) workload accelerators  (e.g., the AMD/Xilinx's ACAP platform \cite{versal}, the Sambanova Reconfigurable Dataflow Architecture \cite{sambanova}, and the Cerebras Wafer Scale Engine \cite{cerebras}). 
These devices have tens to thousands of Processing Elements organized in a 2D grid, communicating using a fast, reconfigurable Network-On-Chip (NoC). Common to all is their amenability to being programmed with a dataflow programming model to favor on-chip data movement and reduce control overheads.
In line with their primary audience, manufacturers of spatial architectures offer integration with popular high-level ML frameworks (e.g., PyTorch and TensorFlow \cite{vitis_ai, sambaflow, cerebras_pytorch}) and release predefined models~\cite{vitis_ai_model_zoo, cerebras_model_zoo}, allowing the users to target these devices for inference tasks conveniently.
%

Despite the promise of massive parallelism and high performance, the scientific community has yet to fully explore the use of such devices in areas other than ML, such as computational science or graph processing. In such cases, programmers have to rely on lower-level APIs (e.g., AMD ADF~\cite{aie-programming-guide}, Cerebras CSL \cite{cerebas_API}) to fully utilize the devices' capabilities. However, the steep learning curve of these APIs and the lack of reusable libraries make it hard for non-experts to explore and leverage these new devices.


In this paper, we present \aieblas, our ongoing effort to design and develop an open-source\footnote{The library is available at: \url{https://github.com/atlarge-research/AIE-BLAS}} implementation of Basic Linear Algebra Routines (\blas) for the AMD AI Engine (AIE), a spatial architecture currently being offered in commodity CPUs~\cite{xdna} and data center accelerators~\cite{versal}.
%
Our goals for the \aieblas library are \textit{1)} to provide ready-to-use numerical routines that can be customized and integrated with other code without requiring the user to write lengthy and complicated lower-level code; \textit{2)} to be easily expandable with new functionalities and optimizations; \textit{3)} to naturally favor on-chip communications using a dataflow approach.

Although our focus in this work is on the AIE architecture, we believe similar design principles and reasoning can also be applied to other spatial architectures.

\begin{figure}
    \centering
    \includegraphics[width=\columnwidth]{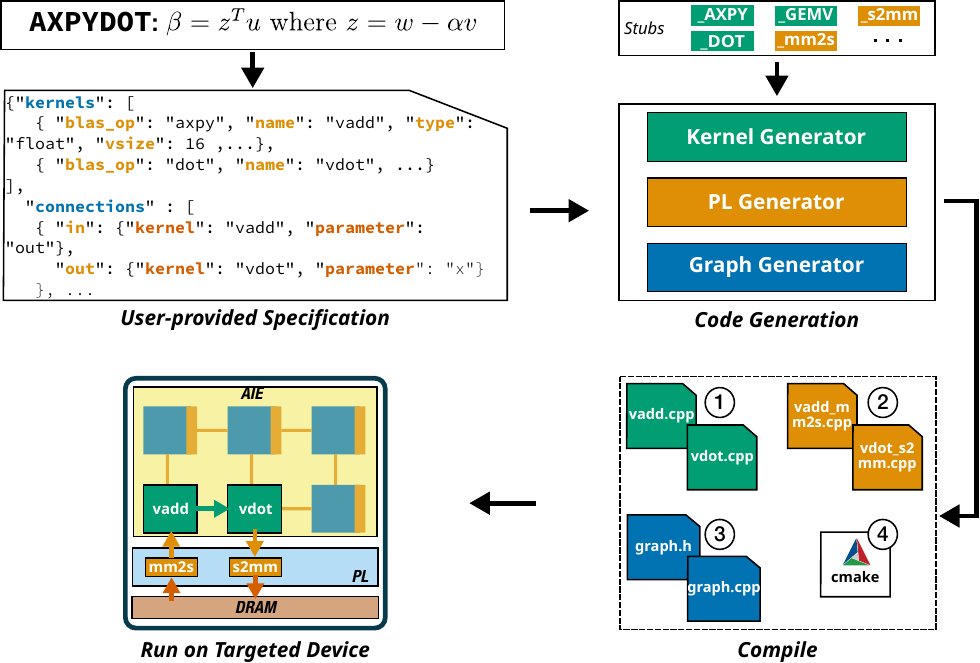}
    \caption{\aieblas development workflow.}
    \label{fig:aie}
\end{figure}


\section{Background}
\begin{figure}[t]
    \centering
    \includegraphics[width=0.7\columnwidth]{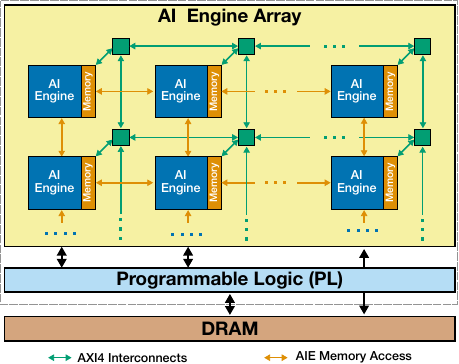}
    \caption{Overview of the AMD Versal ACAP Architecture.}
    \label{fig:acap}
\end{figure}

\aieblas targets Versal Adaptive Compute Acceleration Platform (ACAP) devices.
\Cref{fig:acap} shows the high-level architecture of the VCK5000 development board~\cite{aie-vck5000}. 
The AIE array is organized in an $8\times 50$ grid of 400 AIEs. Each AIE contains a Very Long Instruction Word vector processor and 32KB of local memory. It can share data with the adjacent AIEs by reading/writing directly from/to their local memory. Non-local communications are implemented via AXI4 Streams. The Programmable Logic (PL) component comprises logic blocks, memory, and digital signal processing units (DSPs) that can be used to implement custom logic in hardware. 
The AIE array and PL communicate via multiple AXI interfaces (312 PL $\rightarrow$ AIEs, and 234 AIEs $\rightarrow$ PL), each operating at 4 GB/s.

The AIEs can be programmed using the Adaptive Dataflow (ADF) API~\cite{aie-programming-guide}. The application is represented by a dataflow graph of kernels scheduled one the AIEs. Kernels exchange data by blocks (\textit{windows}) or element by element (\textit{streams}), using the underlying NoC and neighbor interfaces. The PL can be programmed using High-Level Synthesis (HLS) or Register Transfer Level (RTL).


\section{Design and implementation}
\Cref{fig:aie} shows the general development flow
with \aieblas. The user specifies the routine characteristics they need in a JSON file, by indicating information about the type of routine and a unique name for the kernel generation. The user can optionally specify also non-functional parameters, such as windows size, that default to predefined values otherwise.
%
%
%
Starting from the JSON high-level specification, \aieblas generates a design consisting of \one the AIE kernels that implement the required BLAS routines, \two the PL kernels to send and receive data from the device DRAM, \three a dataflow graph to execute the program and, if applicable, connect the AIE kernels as specified, and \four a CMake project to build the design.
Different template-based code generators are in charge of producing the code for the various design components.  All can be conveniently extended to implement new functionalities (e.g., a new routine) or improve existing ones (e.g., an optimized implementation of a given routine).


\vspace{.5ex} 

\aieblas routines accept and produce scalar data using \textit{streams}. For vectors and matrices, we let routines accept and produce \textit{windows}. This approach has several benefits. First, windows 
are stored on the local memory, and  
can be accessed using a wider datapath compared to AXI4 streaming interconnects, allowing us to fully leverage the AIE vector processor. Second, they allow decoupling communications between two communicating AIEs, which is useful for on-chip communications.
Kernel code is vectorized to fully utilize the computing capabilities of the AIEs. The user can set the vector width in the JSON specification, which defaults to the maximum supported (512 bits).
%

\vspace{.5ex} 

Numerical computations can be composed of two or more routines that share data. For instance, the example of \Cref{fig:aie} computes an \axpydot ($\beta = z^Tu$ with $z = w - \alpha v$, where $w$, $v$, and $u$ are vectors,  and $\alpha$ and $\beta$ are scalars, \cite{extended_blas}). This can be implemented by first performing  a vector addition (\axpy), and then using its output as the input of the subsequent \texttt{dot} product. Rather than exchanging data via off-chip memory, we want to favor on-chip communications, composing the routines in a dataflow graph. In this way, we reduce the amount of expensive off-chip accesses and allow for the pipeline executions of multiple routines. AIEBLAS gives users the option to specify connections between BLAS routines in the JSON specification, and the code generator will produce the corresponding dataflow graph definition. If a routine input/output is not connected to another routine, \aieblas will create a PL kernel to load/store the data from off-chip memory.
%
By default, \aieblas relies on the AIE compiler for the kernel placements. However, for larger designs, it may be necessary to provide placement hints to the compiler to generate a floorplan in a reasonable time. To accomplish this, users can set an optional field in the JSON configuration specifying a placement constraint for each kernel.

\section {Initial results}
We evaluated the current implementation of \aieblas on an AMD VCK5000.
The code has been compiled with AMD Vitis v2022.2 and GCC 11.4.0. The host has two 10-cores Intel Xeon Silver 4210R operating at 2.4GHz (no Hyper-Threading), and 256 GB of DDR4 memory. 
For the CPU benchmarks we use OpenBLAS 0.3.27, and optimization flag \texttt{-O3}.
%
\begin{figure}
    \centering
    \includegraphics[width=0.9\columnwidth]{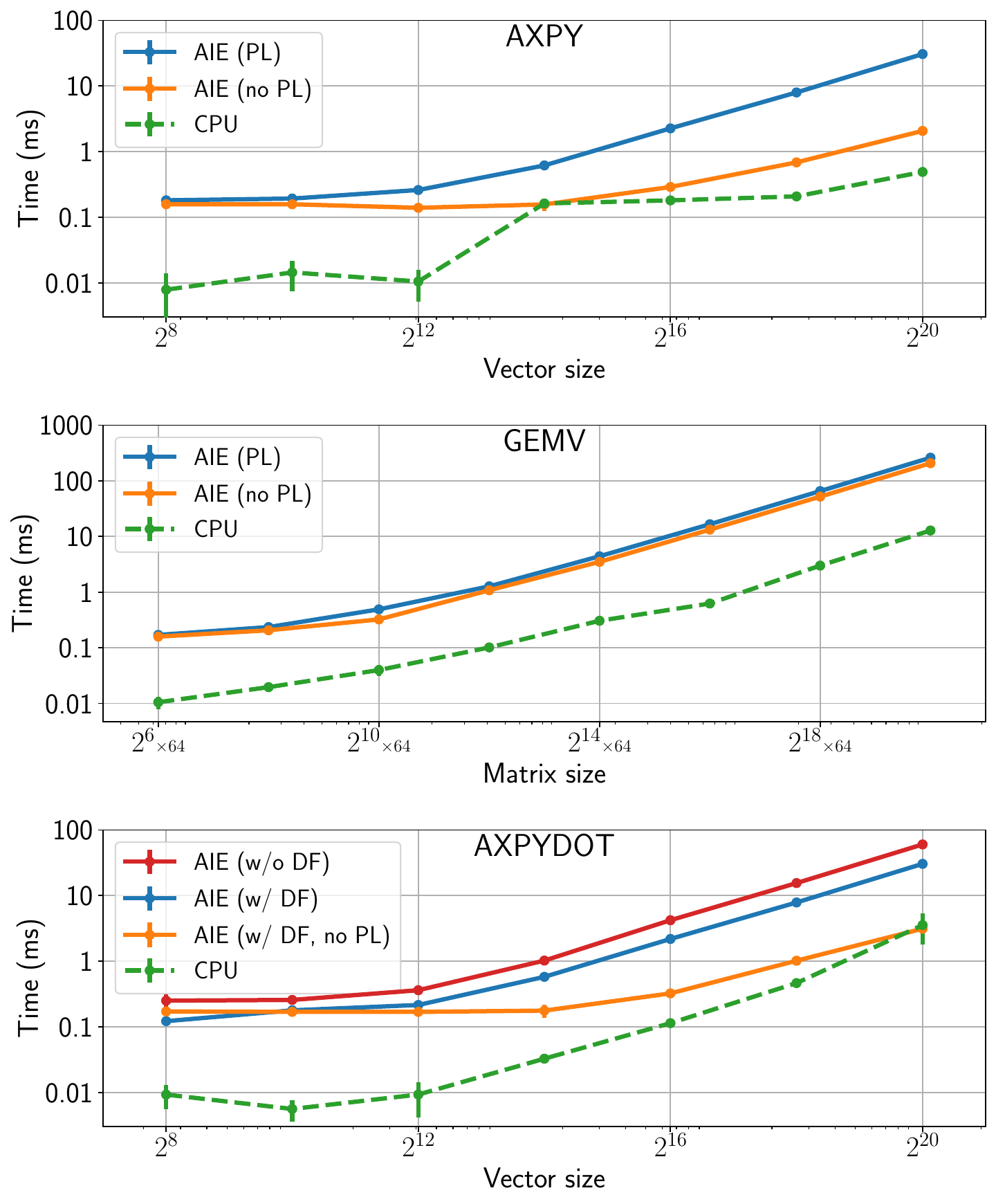}
    \caption{\aieblas evaluation results for different input sizes. We considered implementation with data stored on off-chip memory and movers in programmable logic (PL), and with data being synthetically generated on the AIE array (no PL). For \axpydot, we considered the dataflow (w/ DF) and no-dataflow (w/o DF) implementations.}
    \label{fig:results}
\end{figure}
We considered the vector addition (\axpy) and matrix-vector multiplication routines (\gemv) as well as the composed \axpydot. \Cref{fig:results} reports the averaged execution times for different input sizes.

For the single routines, we tested an implementation that uses PL kernels to read/write data from DRAM and an implementation where the data is generated directly on-chip. The latter results in reduced running time, highlighting the impact of off-chip access on the performance of memory-bound computations. This emphasizes the need to optimize off-chip memory reads (e.g., via burst transfers) and to consider using multiple AXI ports and leverage the various AIE-PL interfaces. 
For the \axpydot routine, we evaluated both dataflow and no-dataflow implementations. As expected, the dataflow approach doubled the performance, indicating that pipelined execution offers significantly better performance. Also in this case, optimizing memory accesses can improve performance. 
Finally, CPU performance is generally better (up 10x) than AIE implementations for all scenarios considered. This is due to OpenBLAS's optimized multicore implementation, an suggests that further optimizations are needed for AIE to achieve competitive performance.


\section{Conclusion and Future Work}
In this paper, we presented the design of \aieblas, a BLAS library for the AMD AIE spatial architecture. The library uses automatic code generation to produce architecture-specific code based on the user’s higher-level specification, significantly increasing user productivity. Initial
performance results demonstrate how dataflow composition is necessary to favor on-chip communications and enable the pipelined execution of multiple routines. The results also highlight the need for more spatial parallelism. Indeed, it is crucial to exploit more parallelism in the PL, to leverage the multiple PL-AIE interfaces and saturate available off-chip memory bandwidth, and via multi-AIE routine implementations, to improve performance even for communication-bound routines.

We intend to continue developing \aieblas in multiple directions. First, we want to improve its performance by \textit{1)} optimizing off-chip memory accesses, \textit{2)} systematically supporting multi-AIEs routines, to exploit the several AIE-PL interfaces and better leverage the available spatial parallelism, and \textit{3)} supporting tiling to reduce off-chip requests further. 
Second, we want to increase BLAS coverage by implementing more routines, also by considering state of the art implementations~\cite{gemm1, gemm2, gemm3}. Finally, by publicly releasing \aieblas, we want to engage the community in developing this and other software
libraries for current and future spatial architectures.

\section*{Acknowledgement}
This work was supported in part by AMD under the Heterogeneous Accelerated Compute Clusters (HACC) program and by the Dutch National Growth Fund through
the 6G FNS project.

\bibliographystyle{IEEEtran}
\bibliography{bibliography}

\begin{thebibliography}{10}
\providecommand{\url}[1]{#1}
\csname url@samestyle\endcsname
\providecommand{\newblock}{\relax}
\providecommand{\bibinfo}[2]{#2}
\providecommand{\BIBentrySTDinterwordspacing}{\spaceskip=0pt\relax}
\providecommand{\BIBentryALTinterwordstretchfactor}{4}
\providecommand{\BIBentryALTinterwordspacing}{\spaceskip=\fontdimen2\font plus
\BIBentryALTinterwordstretchfactor\fontdimen3\font minus \fontdimen4\font\relax}
\providecommand{\BIBforeignlanguage}[2]{{%
\expandafter\ifx\csname l@#1\endcsname\relax
\typeout{** WARNING: IEEEtran.bst: No hyphenation pattern has been}%
\typeout{** loaded for the language `#1'. Using the pattern for}%
\typeout{** the default language instead.}%
\else
\language=\csname l@#1\endcsname
\fi
#2}}
\providecommand{\BIBdecl}{\relax}
\BIBdecl

\bibitem{versal}
\BIBentryALTinterwordspacing
B.~Gaide, D.~Gaitonde, C.~Ravishankar, and T.~Bauer, ``Xilinx adaptive compute acceleration platform: Versaltm architecture,'' in \emph{Proceedings of the 2019 ACM/SIGDA International Symposium on Field-Programmable Gate Arrays}, ser. FPGA '19.\hskip 1em plus 0.5em minus 0.4em\relax New York, NY, USA: Association for Computing Machinery, 2019, p. 84–93. [Online]. Available: \url{https://doi.org/10.1145/3289602.3293906}
\BIBentrySTDinterwordspacing

\bibitem{sambanova}
M.~Emani, V.~Vishwanath, C.~Adams, M.~E. Papka, R.~Stevens, L.~Florescu, S.~Jairath, W.~Liu, T.~Nama, and A.~Sujeeth, ``Accelerating scientific applications with sambanova reconfigurable dataflow architecture,'' \emph{Computing in Science \& Engineering}, vol.~23, no.~2, pp. 114--119, 2021.

\bibitem{cerebras}
S.~Lie, ``Multi-million core, multi-wafer ai cluster,'' in \emph{2021 IEEE Hot Chips 33 Symposium (HCS)}, 2021, pp. 1--41.

\bibitem{vitis_ai}
``{AMD Vitis AI},'' \url{https://www.xilinx.com/products/design-tools/vitis/vitis-ai.html}.

\bibitem{sambaflow}
Sambanova, ``{Accelerated Computing with a Reconfigurable Dataflow Architecture},'' \url{https://sambanova.ai/hubfs/23945802/SambaNova_Accelerated-Computing-with-a-Reconfigurable-Dataflow-Architecture_Whitepaper_English-1.pdf}, 2022.

\bibitem{cerebras_pytorch}
``{Supporting PyTorch on the Cerebras Wafer Scale Engine},'' \url{https://www.cerebras.net/blog/supporting-pytorch-on-the-cerebras-wafer-scale-engine/}.

\bibitem{vitis_ai_model_zoo}
``{Vitis AI Model Zoo},'' \url{https://github.com/Xilinx/Vitis-AI/tree/master/model_zoo}.

\bibitem{cerebras_model_zoo}
``{Cerebras Model Zoo},'' \url{https://github.com/Cerebras/modelzoo/}.

\bibitem{aie-programming-guide}
\BIBentryALTinterwordspacing
\emph{AI Engine Kernel and Graph Programming Guide (UG1079)}, {Advanced Micro Devices, Inc.} [Online]. Available: \url{https://docs.amd.com/r/2022.2-English/ug1079-ai-engine-kernel-coding}
\BIBentrySTDinterwordspacing

\bibitem{cerebas_API}
J.~Selig, ``{The Cerebras Software Development Kit: A Technical Overview },'' Cerebras, Tech. Rep., 2022.

\bibitem{xdna}
A.~Rico, S.~Pareek, J.~Cabezas, D.~Clarke, B.~Ozgul, F.~Barat, Y.~Fu, S.~Münz, D.~Stuart, P.~Schlangen, P.~Duarte, S.~Date, I.~Paul, J.~Weng, S.~Santan, V.~Kathail, A.~Sirasao, and J.~Noguera, ``Amd xdna™ npu in ryzen™ ai processors,'' \emph{IEEE Micro}, pp. 1--10, 2024.

\bibitem{aie-vck5000}
\BIBentryALTinterwordspacing
\emph{VCK5000 Versal Development Card}, {Advanced Micro Devices, Inc.} [Online]. Available: \url{https://www.xilinx.com/products/boards-and-kits/vck5000.html}
\BIBentrySTDinterwordspacing

\bibitem{extended_blas}
S.~Blackford, J.~Demmel, J.~Dongarra, I.~Duff, S.~Hammarling, G.~Henry, M.~Heroux, L.~Kaufman, A.~Lumsdaine, A.~Petitet, R.~Pozo, K.~Remington, and C.~Whaley, ``An updated set of basic linear algebra subprograms ({BLAS}),'' \emph{ACM Trans. Math. Softw.}, vol.~28, no.~2, pp. 135--151, Jun. 2002.

\bibitem{gemm1}
\BIBentryALTinterwordspacing
E.~Taka, D.~Gourounas, A.~Gerstlauer, D.~Marculescu, and A.~Arora, ``Efficient approaches for gemm acceleration on leading ai-optimized fpgas,'' 2024. [Online]. Available: \url{https://arxiv.org/abs/2404.11066}
\BIBentrySTDinterwordspacing

\bibitem{gemm2}
Z.~Wu, M.~Gokhale, S.~Lloyd, and H.~Patel, ``Sccl: An open-source systemc to rtl translator,'' in \emph{2023 IEEE 31st Annual International Symposium on Field-Programmable Custom Computing Machines (FCCM)}, 2023, pp. 23--33.

\bibitem{gemm3}
J.~Zhuang, Z.~Yang, and P.~Zhou, ``High performance, low power matrix multiply design on acap: from architecture, design challenges and dse perspectives,'' in \emph{2023 60th ACM/IEEE Design Automation Conference (DAC)}, 2023, pp. 1--6.

\end{thebibliography}

\end{document}